\documentstyle[12pt]{article}
\begin{document}
\begin{flushright} \small August, 1995 \\McGill/95-47 \\ hep-th/9508179
\end{flushright}
\vspace{1.2em}
\begin{center} {\LARGE {\bf {Chiral bosonization as a Duality}}}
\end{center}
\vspace{1.2em}
\begin{center} {\large {{\sc Mohammad R. Garousi}}}
\end{center}

\begin{center}
{\large { Department of Physics, McGill University \\
        3600 University St. Montreal, T3A 2T8 \\
                Canada}}
\end{center}

\vspace{3.2em}

\abstract{%
We demonstrate that the technique of abelian bosonization through duality transformations can be extended to gauging anomalous symmetries. The example of a Dirac fermion theory is first illustrated. This idea is then also applied to bosonize a  chiral  fermion by gauging its chiral phase symmetry.
}
\\ \\ \\

Two dimensional field theory has long  proven to be a fruitful area of investigation. An interesting phenomenon in two dimensional field theory known to physicists for some time is bosonization [1]. It states the equality of the correlation function of some operators of two apparently different theories, one being a theory of a  Dirac fermion and the other that  of a scalar boson. Conventionally this equality is expressed in  terms of bosonization rules [2][3].  A very elaborate framework for bosonization is smooth bosonization [4]. In this framework one deals with a family of theories containing both fermion and boson fields in which one may smoothly interpolate between a pure fermionic theory and a  pure bosonic one. So considering  only the bosonic and fermionic theories limits the discussion to  a small sector of this larger framework. In a related point of view,  C.P. Burgess and F. Quevedo [5] have shown  how to systematically derive the  bosonization rules in two dimensions  as a particular case of a duality transformation [6]. The technique is, briefly stated, that first the phase symmetry of the fermion theory is gauged which  adds a new gauge field $  A_{\mu}$ to the theory. The corresponding field strength $F_{\mu \nu}$ is then constrained so that it  vanishes everywhere. This constraint can be incorporated through a  Lagrange multiplier field $\phi$. To obtain the bosonized action one  fixes the gauge ($\partial \cdot A=0$)  and  integrates out the original fermion field as well as  the new  gauge field $A_{\mu}$. One is then left with a bosonic action in terms of the Lagrange multiplier $\phi$.  On the other hand, integrating over the Lagrange multiplier field $\phi$ and the gauge field $A_{\mu}$  recovers the original fermionic theory. In other words, this method is a change of variable with the  path integral, a change from a  fermionic  variable to a  bosonic  variable.

A slightly different way of representing the same  idea is with the following path integral identity\footnote{Throughout the paper space-time is assumed to be flat with topology $ {\bf {R}}_{2}$. We also ignore all \\
 irrelevant normalization constants in path integral.}:
\begin{equation}
\int D \psi D A_{\mu} \Delta [A_{\mu}]\exp [i \hat{S}_{F}[\psi,A_{\mu},a_{\mu},b_{\mu}]]=\int D \psi\exp [i S_{F}[\psi,a_{\mu},b_{\mu}]]
\end{equation}
where $S_{F}[\psi,a_{\mu},b_{\mu}]$ is an action for fermionic theory and $a_{\mu}$ and $b_{\mu}$ are external fields which couple to $\psi$ (see below) and we assumed $\hat{S}_{F}[\psi,A_{\mu}=0,a_{\mu},b_{\mu}]=S_{F}[\psi,a_{\mu},b_{\mu}]$. Here $\Delta [A_{\mu}(x,t)]=\prod _{x,t} \delta [A_{\mu}(x,t)]$ is the functional delta function. Eq.(1) shows that one has a great deal of arbitrariness in adding  $A_{\mu}$ to the action provided  one takes the functional integral over $D A_{\mu} \Delta [A_{\mu}]$. For topologically trivial space-time one can write $\Delta [A_{\mu}]$ as $\Delta [\partial \cdot A]\Delta [\frac {1}{2} \epsilon^{\mu \nu} F_{\mu \nu}]$ and the bosonization  can be carried out by  writing one of these functional delta functions as its functional Fourier transform and performing the $\psi$ and $A_{\mu}$ integrations. The Fourier fields then play the role of new variables. Following this prescription, in general, leads  to a  nonlocal theory  unless $A_{\mu}$ is introduced in to $\hat{S}_{F}$ in a special way. In [5] it is shown that if one adds $A_{\mu}$ through the gauging phase symmetry of the original Dirac fermionic theory, one ends up with an action in terms of the Lagrange multiplier implementing  $\Delta [\frac {1}{2} \epsilon^{\mu \nu} F_{\mu \nu}]$  which is the bosonized theory  and local as well. In this letter we will show that if one adds $A_{\mu}$ through gauging the chiral phase symmetry of the original Dirac fermion action, one  ends up with the same local theory. We will also demonstrate that the new bosonic field is a Lagrange multiplier implementing either  $\Delta [\frac {1}{2} \epsilon^{\mu \nu} F_{\mu \nu}]$ or  $\Delta[\partial \cdot A]$ in general. In certain special cases though there is no freedom in this choice, for example for non-chiral rotation studied in [5] one must use $\Delta[\frac {1}{2} \epsilon^{\mu \nu} F_{\mu \nu}]$. Finally, we  apply  this idea, to illustrate how to bosonize chiral fermionic theory by adding $A_{\mu}$ into the theory through gauging its chiral phase symmetry.\\ \\

We start with the following generating functional for massless Dirac fermion:
\begin{equation}
Z[a_{\mu},b_{\mu}]=\int D \psi \exp [i S_{F}[\psi ,a_{\mu},b_{\mu}]]
\end{equation}
where
\begin{equation}
S_{F}[\psi , a_{\mu},b_{\mu}]=\int d^{2}x(-\overline{\psi}\gamma^{\mu}\partial_{\mu}\psi + i \overline{\psi}\gamma^{\mu}\psi a_{\mu}+i \overline{\psi}\gamma^{\mu}\gamma_{3}\psi b_{\mu}). 
\end{equation}
and $a_{\mu}$ and $b_{\mu}$ are external fields.
This classical action has the following chiral phase symmetry \footnote{Our conventions are: $ x^{0}=t, x^{1}=x, \eta^{11}=-\eta^{00}=\epsilon^{01}=1, \gamma_{0}=i \sigma_{1}, \gamma_{1}=\sigma_{2}, \gamma_{3}\equiv \gamma^{0}\gamma^{1}= \sigma_{3}$ and \\
 $\gamma_{L}=\frac {1}{2}(1+\gamma_{3})$.}, 
\begin{equation}
\psi \longrightarrow \exp[\frac {i}{2}(q+1)\alpha +\frac{i}{2}(q-1)\alpha \gamma_{3}]\psi
\end {equation}
which is a constant phase rotation of right (left) hand component of $\psi$ by $\alpha$ ($q \alpha$). In the following, 
we gauge this symmetry and constrain the gauge field $A_{\mu}$ so that it vanishes. Following the above discussion  the generating functional can be written as,
\begin{equation}
Z[a_{\mu},b_{\mu}]=\int D \psi D A_{\mu}\Delta [\partial \cdot A]\Delta
[\frac {1}{2} \epsilon^{\mu \nu} F_{\mu \nu}] \exp[i S[\psi,A_{\mu},a_{\mu},b_{\mu}]] 
\end{equation}
where
\begin{equation}
S[\psi,A_{\mu},a_{\mu},b_{\mu}]=\int d^{2}x[\overline{\psi}\gamma^{\mu}(-\partial _{\mu}+\frac {i}{2}(q+1)A_{\mu}+\frac{i}{2}(q-1)\gamma_{3}A_{\mu}+i a_{\mu}+i \gamma_{3}b_{\mu})\psi].
\end{equation}
Using the identity $\gamma^{\mu}\gamma_{3}=\gamma_{\nu}\epsilon ^{\nu \mu}$, one can perform the fermionic integral in eq.(5). The result is [6]:
\begin{equation}
\int D \psi \exp [i \int d^{2}x [\overline{\psi}\gamma^{\mu}(-\partial_{\mu}
 +i \hat{A}_{\mu})\psi ]]=\exp [- \frac {i}{8 \pi}\int d^{2}x \hat{F}_{\mu \nu}\epsilon^{\mu \nu} \Box ^{-1}\hat{F}_{\mu \nu}\epsilon^{\mu \nu}].
\end{equation}
where
\begin{equation}
\hat{A}_{\mu} \equiv \frac {1}{2}(q+1)A_{\mu}+\frac {1}{2}(q-1)\epsilon_{\mu \nu}
A^{\nu}+a_{\mu}+\epsilon_{\mu \nu} b^{\nu}.
\end{equation}

Now at this point we have to add a new bosonic field to the theory through  Fourier transform of one of the  delta functionals. We may write the following functional Fourier transform:
\begin{equation}
\Delta [\frac {1}{2} \epsilon^{\mu \nu} F_{\mu \nu}]=\int D \phi\exp[i\int d^{2}x [\frac {1}{2}\phi\, \epsilon^{\mu \nu} F_{\mu \nu}]].
\end{equation}
Replacing eq.(7) and eq.(9) into eq.(5), the generating functional becomes
\begin{equation}
Z[a_{\mu},b_{\mu}]=\int D \phi D A_{\mu}\Delta [\partial \cdot A] \exp[i S [\phi ,A_{\mu},a_{\mu},b_{\mu}]]
\end {equation}
where
\begin{equation}
S [\phi ,A_{\mu},a,b]=\int d^{2}x[\frac{-1}{2 \pi}(\frac{1}{4}(q+1)\epsilon^{\mu \nu} F_{\mu \nu}+B)\Box^{-1}(\frac{1}{4}(q+1)\epsilon^{\mu\nu} F_{\mu \nu}+B)+\frac {1}{2}\phi \epsilon^{\mu \nu} F_{\mu \nu}]
\end {equation}
and  we defined $ B \equiv \epsilon ^{\mu
 \nu}\partial_{\mu} a_{\nu} +\partial \cdot b $ and set $\partial \cdot A=0$. Now we finally perform the integration over $D A_{\mu}$. From $ \partial \cdot A =0 $ we conclude that $ A _ {\mu}=  \epsilon _ {\mu \nu} \partial ^ {\nu} \lambda $, so we do change of variables as $ A_{\mu} \longrightarrow \lambda $. Since determinant of ordinary derivative is a constant, the Jacobian for this transformation is an irrelevant constant, so $ D A_{\mu} \Delta [\partial  \cdot
A] = D \lambda $. In terms of $ \lambda $ we also have $ 
F_{\mu
 \nu}\epsilon ^ {\mu
\nu} = 2 \Box \lambda $. Therefore, equation (10) can be written as\footnote{We assume our fields to fall to zero at spatial infinity sufficiently quickly to permit the neglect \\
 of all surface terms.},
\begin{equation}
Z[a_{\mu},b_{\mu}]=\int D \phi D {\lambda}
\exp[i \int d^{2}x [-\frac {1}{2 \pi } B \Box ^{-1} B+Q(\lambda)]]
\end{equation}
where 
\begin{equation}
Q(\lambda)=- \frac {1}{8 \pi}(q+1)^{2}\lambda \Box \lambda +[- \frac {1}{2 \pi}(q+1) B + \Box \phi ] \lambda.
\end{equation}
Since $ Q(\lambda)$ is quadratic, it can be expanded around its saddle point, $\lambda_{c}$, as
\begin{equation}
Q(\lambda)=- \frac {1}{8 \pi}(q+1)^{2}(\lambda - \lambda _{c})\Box (\lambda - \lambda _{c}) + Q(\lambda _{c})
\end{equation}
where the saddle point satisfies the following equation
\begin{equation}
- \frac {1}{4 \pi}(q+1)^{2} \Box \lambda_{c} +(- \frac {1}{2 \pi}(
q+1) B + \Box \phi ) =0.
\end{equation}
The resulting $\lambda - \lambda _{c}$ integration is an irrelevant constant. Therefore, equation (12) reduces to
\begin{equation}
Z[a_{\mu},b_{\mu}]=\int D \phi \exp[i \int d^{2}x [-\frac {1}{2 \pi } B \Box ^{-1} B+Q(\lambda _{c} )]]
\end{equation}
where $Q(\lambda _{c})$ simplifies to
\begin{equation}
Q(\lambda _{c})= \frac {2 \pi } { (q+1)^{2} }\phi \Box \phi - \frac {2}{q+1} \phi B + \frac {1}{2 \pi } B \Box ^{-1} B.
\end{equation}
Replacing this into equation (16) and using the definition for $ B $, we end up with following  bosonic action:
\begin{equation}
S_{B}[\phi , a_{\mu},b_{\mu}]= \int d^{2}x [-\frac {2 \pi }{(q+1)^{2}} \partial_{\mu} \phi \partial ^ {\mu}\phi + \frac {2}{q+1}\partial_{\mu}\phi ( \epsilon ^ {\mu
\nu} a _ {\nu} + b ^ {\mu} ) ].
\end{equation}
Note that the nonlocal terms from  performing the $\psi$ and $A_{\mu}$ integrations  are  canceled as in the non-chiral case [5]. In terms of the canonically-normalized scalar variable, $ \Lambda = \frac {2 \sqrt {\pi }}{q+1} \phi $, the dual action takes its standard form:
\begin{equation}
S_{B}[\Lambda,a_{\mu},b_{\mu}] = \int d^{2}x [-\frac {1}{2} \partial_{\mu} \Lambda \partial ^ {\mu}\Lambda + \frac {1}{\sqrt{\pi}} \partial_{\mu}\Lambda b^ {\mu} + \frac {1}{\sqrt{\pi}} \partial_{\mu}\Lambda  \epsilon ^ {\mu
\nu} a _ {\nu}].
\end{equation}
Comparing the coefficients of $a_{\mu}$ and $b_{\mu}$ in equations (3) and (19) shows that the currents in these two theories are related by
\begin{equation}
i \overline{\psi}\gamma^{\mu}\psi \longleftrightarrow - \frac {1}{\sqrt{\pi}}\epsilon ^ {\mu \nu}\partial_{\nu}\Lambda \;\;\; , \;\;\; i \overline{\psi}\gamma^{\mu}\gamma_{3}\psi \longleftrightarrow\frac {1}{\sqrt{\pi}}\partial^{\mu}\Lambda.
\end{equation}
For $ q=-1$ eq.(14) becomes linear and eventually eq.(19) turns out to be nonlocal. To cure this problem we use the following  functional Fourier transform instead of eq.(9)
\begin{equation}
\Delta [\partial \cdot A ]= \int D \phi \exp [i \int d ^{2} x [\phi \partial \cdot A ] ].
\end{equation}
Replacing eq.(7) and eq.(21) into eq.(5), the generating functional becomes
\begin{equation}
Z[a_{\mu},b_{\mu}]= \int D \phi D A_{\mu} \Delta [\frac {1}{2}\epsilon ^ {\mu \nu } F_{\mu \nu }]\exp [ i S [\phi , A_{\mu} ,a_{\mu},b_{\mu}]]
\end{equation}
where
\begin{equation}
S [\phi , A,a_{\mu},b_{\mu}]=\int d^{2} x [- \frac {1}{2 \pi}(\frac {1}{2}(q-1)\partial \cdot A+B) \Box ^ {-1} (\frac {1}{2}(q-1)\partial \cdot 
A+B) + \phi \partial \cdot 
A ]
\end{equation}
and we set $\epsilon^{\mu \nu}F_{\mu \nu}=0$. To perform the integral over the gauge potential, $ A_{\mu}$ , we do change of variables as $A_{\mu} \longrightarrow \rho $ with $ A_{\mu} \equiv \partial  _ {\mu } \rho $. By doing the same steps as before, we end up with the following bosonic action: 
\begin{equation}
S_{B} [\phi ,a,b]=\int d ^{2} x [- \frac {2 \pi}{(q-1)^{2}} \partial _{\mu} \phi \partial ^ {\mu}\phi +\frac {2}{q-1} \partial _{\mu}\phi (\epsilon ^ {\mu \nu } a _ {\nu} +b ^ {\mu} )]
\end{equation}
which turns out to be nonlocal for $q=1$. In terms of the canonically-normalized scalar variable, $ \Lambda =\frac {2 \sqrt{\pi}}{q-1} \phi $, the dual action is equal to equation (19). This ultimately shows the use of anomalous symmetry in duality transformations for  bosonizatoin of  Dirac massless fermion theory in (1+1) dimension. However, we emphasize that the bosonic field in the dual theory must be chosen to be the Lagrange multiplier implementing  $\Delta[\frac {1}{2}\epsilon^{\mu \nu}F_{\mu \nu}]$ in the case  $q=1$ (non-chiral phase rotation), implementing  $\Delta[\partial \cdot A]$ in the case  $q=-1$ and  implementing  either of them in other cases.\\
 
Now using the fact that anomalous symmetries potentially has the ability to bosonize Dirac fermion field, we shall bosonize chiral fermion field by employing its anomalous chiral phase symmetry. Consider the following generating functional for chiral fermion particles:
\begin{equation}
Z[a_{\mu}]=\int D \psi _{R}\exp[i \int d^{2}x \overline{\psi}_{R}\gamma^{\mu}(-\partial_{\mu}+i a_{\mu})\psi_{R}]
\end{equation}
where $a_{\mu}$ is an external field. This classical action has phase symmetry, $\psi_{R}\longrightarrow \exp[i \alpha ]\psi_{R}$, where $\alpha$ is constant. Adding an $A_{\mu}$ field through gauging this symmetry and constraining the gauge field to be zero everywhere, one can write eq.(25) for trivial space-time topology as,
\begin{equation}
Z[a_{\mu}]=\int D \psi_{R} D A_{\mu}\Delta [\partial \cdot A]\Delta
[\frac {1}{2} \epsilon^{\mu \nu} F_{\mu \nu}] \exp[i S_{R}[\psi_{R},A_{\mu},a_{\mu}]]
\end{equation}
where
\begin{equation}
S_{R}[\psi_{R},A_{\mu},a_{\mu}]=\int d^{2}x \overline{\psi}_{R}\gamma^{\mu}(-\partial_{\mu}+i A_{\mu}+i a_{\mu})\psi_{R}. 
\end{equation}
The result of performing $D \psi _{R}$ integral is [8]:
\begin{equation}
\int D \psi_{R}\exp[i \int d^{2}x[ \overline{\psi}_{R}\gamma^{\mu}(-\partial_{\mu}+i \hat{A}_{\mu})\psi_{R}]]=\exp[\frac{-i}{8 \pi}\int d^{2}x \hat{F}_{\mu \nu}
\epsilon ^ {\mu \nu} \Box ^{-1}(\hat{F}_{\mu \nu}\epsilon ^ {\mu \nu}-2\partial \cdot \hat{A})]
\end{equation}
where $\hat{A}_{\mu}\equiv A_{\mu}+a_{\mu}$. Using the identity $\epsilon^{\alpha \beta}\epsilon^{\mu \nu}=\eta ^{\alpha \nu}\eta^{\beta \mu}-\eta^{\alpha \mu}\eta^{\beta \nu}$, one can show that $(\hat{F}_{\mu \nu}\epsilon ^ {\mu \nu}+2\partial \cdot \hat{A})\Box^{-1}(\hat{F}_{\mu \nu}\epsilon ^ {\mu \nu}-2\partial \cdot \hat{A})=4\hat{A} \cdot\hat{ A}$. Therefore, eq.(28) can be written as
\begin{equation}
L.H.S.= \exp[\frac{-i}{16 \pi}\int d^{2}x( \hat{F}_{\mu \nu}
\epsilon ^ {\mu \nu}-2\partial \cdot\hat{A}) \Box ^{-1}(\hat{F}_{\mu \nu}\epsilon ^ {\mu \nu}-2\partial 
\cdot \hat{A})-\frac {i}{4 \pi}\int d^{2}x \hat{A}\cdot \hat{A}].
\end{equation}
Using the fact that eq.(28) is fixed only up to a local polynomial in the field $\hat{A}_{\mu}$ and its derivatives \footnote {The same freedom occurs  in performing the fermion path integral in eq.(7).} the last term in eq.(29) can be dropped by adding a local counterterm.

 Now we have to add a new bosonic field in to the theory through Fourier transform of one of the delta functions in eq.(26). Since phase symmetry in this case in corresponding to $q=0$ of previous example, the new scalar field can be Lagrange multiplier of either of the constraints in eq.(26).  Writing $\Delta [\partial \cdot A]$ with a  Fourier transform, eq.(26) becomes
\begin{equation}
Z[a_{\mu}]=\int D \phi D A_{\mu} \Delta
[\frac {1}{2} \epsilon^{\mu \nu} F_{\mu \nu}]\exp[i S[A_{\mu},a_{\mu}]]
\end{equation}
where
\begin{equation}
S[A_{\mu},a_{\mu}]=\int d^{2}x [\frac{-1}{4 \pi}(-\partial \cdot A +\epsilon ^{\mu \nu} \partial_{\mu} a_{\nu}-\partial \cdot a)\Box^{-1}(-\partial \cdot A +\epsilon ^{
\mu \nu} \partial_{\mu} a_{\nu}-\partial \cdot a)+\phi \partial \cdot A]
\end{equation}
and we set $\epsilon^{\mu \nu}F_{\mu \nu}=0$. Performing the $D A_{\mu}$ integral, one ends up with the following local bosonic action:
\begin{equation}
S_{RB}[\phi,a_{\mu}]=\int d^{2}x [-\pi\partial _{\mu} \phi \partial^{\mu} \phi-(\partial _{\mu} \phi-\partial^{\nu}\phi\epsilon_{\nu \mu})a^{\mu}].
\end{equation}
In terms of canonically-normalized scalar variable, $\Lambda =- \sqrt{2 \pi}\phi$, the bosonized chiral fermion action is
\begin{equation}
S_{RB}[\Lambda,a_{\mu}]=\int d^{2}x [-\frac{1}{2} \partial_{\mu}\Lambda\partial^{\mu}\Lambda+\frac{1}{ \sqrt{2 \pi}}(\partial_{\mu}\Lambda -\partial^{\nu}\Lambda \epsilon_{\nu \mu})a^{\mu}].
\end{equation}
Comparing the coefficient $a_{\mu}$ in eq.(33) and eq.(25) shows that the current in these two theories are related by
\begin{equation}
i\overline{\psi}_{R}\gamma_{\mu}\psi _{R}=i \overline{\psi}\gamma_{\mu}\frac{1-\gamma_{3}}{2}\psi \longleftrightarrow \frac{1}{\sqrt{2 \pi}}(\partial_{\mu}\Lambda+\epsilon_{\mu \nu}\partial^{\nu}\Lambda).
\end{equation}
To compare eq.(33) with original chiral fermion action, we  write both action in  light cone coordinate, $x^{\pm}=\frac{1}{\sqrt{2}}(x \pm t)$. The chiral fermion action takes the following form:
\begin{equation}
S_{F}[\psi_{R},a_{\mu}]=\int d^{2}x \psi_{R}^{*}(-\partial _{+} +i a_{+})\psi _{R}
\end{equation}
and using the fact that $\eta _{+-}=1$ and $\epsilon_{+-}=-1$,  eq.(33) becomes
\begin{equation}
S_{RB}[\Lambda,a_{\mu}]=\int d^{2}x (-\frac{1}{2} \partial_{+}\Lambda\partial_{-}\Lambda +\frac{1}{\sqrt{\pi}}\partial_{-} \Lambda a_{+}).
\end{equation}
Eq.(35) shows that $\psi_{R}$ has only right-moving  modes, i.e., $\psi_{R}(x,t)=\psi_{R}(x_{-})$,  and only the $a_{+}$ component couples to it. On the other hand  eq.(36) indicates that $\Lambda$ will have  both right- and left-moving  modes, i.e., $\Lambda (x,t)=\Lambda (x_{+})+\Lambda (x_{-})$, but source only couples to the right-moving  modes through  $\partial_{-}\Lambda (x_{-})$. To compare from symmetry point of view, we add a local counterterm, $\frac {i}{\pi}a_{\mu}a^{\mu}$, to the eq.(31) and it changes eq.(33) to the following form:
\begin{equation}
S_{RB}[\Lambda,a_{\mu}]=\int d^{2}x [-\frac{1}{2}(\partial_{\mu}\Lambda -\frac{1}{\sqrt{2 \pi}}a_{\mu})(\partial^{\mu}\Lambda -\frac{1}{\sqrt
{2 \pi}}a^{\mu})+\frac{1}{2 \sqrt{2 \pi}}\Lambda \epsilon_{\mu \nu}F^{\mu \nu}_{a}]
\end{equation}
where $F^{\mu \nu}_{a}=\partial^{\mu}a^{\nu}-\partial^{\nu}a^{\mu}$. Under $\delta a_{\mu}=\partial_{\mu}\alpha $ and $\delta \Lambda=\frac{1}{\sqrt{2 \pi}}\alpha $ transformation, the bosonic action has the following ``classical'' anomaly:
\begin{equation}
\delta S_{RB}[\Lambda,a_{\mu}]=\frac{1}{4\pi}\int d^{2}x\alpha \epsilon_{\mu \nu}F^{\mu \nu}_{a}
\end{equation}
which reproduces the  quantum chiral anomaly of  fermion theory [9]. This  illustrates our claim about abelian chiral bosonization as a duality transformation.\\ \\ \\ \\ \\ 
{\bf { Discussion}} \\ 

 We have shown how to bosonize massless Dirac fermionic theory  through gauging its chiral phase symmetry. The gauging of the chiral symmetry results in the addition of an $A_{\mu}$ field which is then  constrained   so that it vanishes everywhere. Also new scalar boson field is introduced into the theory by  writing part of $\Delta[A_{\mu}]$ as functional Fourier transform. This  new scalar field is a  Lagrange multiplier implementing  $\Delta[\partial \cdot A]$ in the case  $q=-1$, implementing  $\Delta[\frac{1}{2}\epsilon^{\mu \nu}F_{\mu \nu}]$ in the case  $q=1$ and one is free to choose  either of them in other cases. Thus one is able to bosonize chiral fermion theory by exploiting chiral phase symmetry. In this example, $q=0$, we were free to choose the scalar field as the Lagrange multiplier of $\Delta[\partial \cdot A]$ or $\Delta[\frac {1}{2}\epsilon^{\mu \nu }F_{\mu \nu }]$. 

We now ask the question of why in the case $q \neq \pm 1$ both constraints appear on the same footing whereas in the case of $q=\pm 1$ they do not. The answer is that, in  $q \neq \pm 1 $ case, if one implements both constraints and integrates out $\psi$ and $A_{\mu}$, one  ends up with a nonlocal action containing  two  indistinguishable  scalar fields, the Lagrange multipliers of $\Delta[\partial \cdot A]$ and $\Delta[\frac {1}{2}\epsilon^{\mu \nu }F_{\mu \nu}]$. The local dual action may  then be obtained by integrating out either one of them. In contrast, in the case $q=1$, after integrating out $\psi$ and $A_{\mu}$, one ends up with a local action in terms of the Lagrange multiplier of $\Delta[\frac {1}{2}\epsilon^{\mu \nu}F_{\mu \nu}]$ only and similarly in the case $q=-1$, in terms of the Lagrange multiplier of $\Delta[\partial \cdot A]$. In each of these  cases the second Lagrange multipler  appears trivially in  $\int D\phi \Delta[\Box \phi]$  and so it can be ignored. Therefore, in the present approach  the bosonized theories are local only when expressed in terms of one scalar field. Using another quantization scheme, a  local bosonic theory is produced with two scalars in [12]. The difference between these two results lies in a different choice of local counterterms in the fermionic path integral, i.e., eqs.(7) and (28). 

New bosonic fields could be introduced in to the theory in a slightly different way through writing $\Delta [A_{\mu}]$ as functional Fourier transform. In this case the Lagrange multiplier is a contravariant vector, $\Omega^{\mu}$. In topologically trivial space-time one can make a  change of variables to $A_{\mu}\equiv \partial_{\mu}A+\epsilon_{\mu \nu}\partial^{\nu}B$ and $\Omega^{\mu}\equiv \partial^{\mu}\omega+\epsilon^{\mu \nu}\partial_{\nu}\phi$ where $A \,,B\, , \omega$ and $\phi$ are scalars.  Integrating out $\psi$, $A$ and $B$, one ends up with the same result as in the previous paragraph for different $q$'s.

Smooth bosonization depends on introduction of a collective field through a local chiral transformation, $\psi (x)=\exp[i \Lambda (x) \gamma _{3}]\chi (x)$. By using a Fadeev-Popov type trick, $\Lambda (x)$ is promoted to a dynamical field and the theory containing both fields becomes gauge invariant under  $\chi \longrightarrow e^{i \alpha(x) \gamma_{3}}\chi$ and $\Lambda(x) \longrightarrow\Lambda (x)-\alpha (x)$ transformations with no anomaly.  In this method the gauge fixing function has one real parameter and the pure bosonic theory may be smoothly interpolated to a  pure fermionic one by changing this  parameter. Similarly  the generating functional presented here in  eq.(5), apart from $\Delta [\partial \cdot A]$, is gauge invariant under $ \psi \longrightarrow e^{i \alpha (x) \gamma _{3}}\psi $ (for $q=-1$) and $A_{\mu} \longrightarrow A_{\mu}+\partial _{\mu} \alpha (x) $  with no anomaly  since the presence of $\Delta [\frac{1}{2} \epsilon ^{\mu \nu } F_{\mu \nu }]$ causes the chiral anomaly to vanish.  The  $\Delta [\partial \cdot A ]$ may be regarded as  the gauge fixing term. The  method of smooth bosonization may be extended to use  $\psi(x)=\exp[\frac {i}{2}(q+1)\Lambda(x)+\frac {i}{2}(q-1)\Lambda(x)\gamma_{3}]\chi(x)$ as the local chiral transformation. Extended in this way, one would have a close  parallel between the smooth bosonization technique and the work presented here.

Our extended work on duality illustrates that a theory can be dualized by  gauging any of its  classical symmetries irrespective of quantum anomalies, as one might expect from smooth bosonization technique. Although the dual of  Dirac massless fermion is always the same for any choice of $q$,  it may be possible to discover  models in which  two dual theories  are  different. We also remark that the result of this paper also shows that in  cases in which the original theories have no quantum symmetries but only classical symmetries such as  chiral fermion theories one can still use these techniques  to find their dual theories. Therefore, it may be interesting to  extend the present example,  abelian chiral bosonization, to the nonabelian case [10][11] which also has only classical symmetry. \\ \\ \\

{\large {\bf{Acknowledgements}}} \\

The author would like to acknowledge helpful conversations with Rob Myers, Cliff Burgess, O. Hamidi-Ravari,  J. C. Breckenridge and S. M. Zebarjad. This research was funded by Ministry of Culture and Higher Education of Iran, N.S.E.R.C. of Canada and F.C.A.R du Quebec.

\end {document}